\documentstyle[psfig,epsf]{mn}
\newcommand{\ltaraw}{$\; \buildrel < \over \sim \;$}
\newcommand{\lta}{\lower.5ex\hbox{\ltaraw}}
\newcommand{\gtaraw}{$\; \buildrel > \over \sim \;$}
\newcommand{\gta}{\lower.5ex\hbox{\gtaraw}}

\newcommand{\kms}{{\rm\,km\,s^{-1}}}

\loadboldmathitalic 
\title [Microlensing of planets]
{Gravitational microlensing of planets:
the influence of planetary phase and caustic orientation\footnotemark[1]}
\author[C. E. Ashton \& G. F. Lewis]
{Ceri E. Ashton$^{1}$ \& Geraint F. Lewis$^{2}$\\
$^{1}$
Dept of Physics and Astronomy, Uni. of Wales College of Cardiff,
PO Box 913, Cardiff, CF2 3YB, U.K \\
Email \tt{AshtonCE@Cardiff.ac.uk}\\
$^{2}$
Anglo-Australian Observatory, P.O. Box 296, Epping, NSW 1710, Australia\\
Email \tt{gfl@aaoepp.aao.gov.au}\\
}
\date{\today}
\begin{document} 
\maketitle 
\begin{abstract}
Recent   studies  have  demonstrated   that  detailed   monitoring  of
gravitational microlensing  events can reveal the  presence of planets
orbiting the microlensed source  stars.  With the potential of probing
planets in  the Galactic Bulge and Magellanic  Clouds, such detections
greatly increase  the volume  over which planets  can be  found.  This
paper expands on  these original studies by considering  the effect of
planetary phase on the form of the resultant microlensing light curve.
It is found that  crescent-like sources can undergo substantially more
magnification than  a uniformly illuminated disk,  the model typically
employed  in  studying  such  planets.   In fact,  such  a  circularly
symmetric model is  found to suffer a minimal  degree of magnification
when compared to the crescent  models.  The degree of magnification is
also a strong function of the planet's orientation with respect to the
microlensing  caustic. The  form of  the magnification  variability is
also  strongly  dependent  on  the  planetary  phase  and  from  which
direction it is  swept by the caustic, providing  further clues to the
geometry of  the planetary  system. As the  amount of  light reflected
from  a planet also  depends on  its phase,  the detection  of extreme
crescent-like planets  requires the  advent of 30-m  class telescopes,
while  light  curves  of  planets  at  more  moderate  phases  can  be
determined with today's 10-m telescopes.
\end{abstract}

\footnotetext[1]{Research undertaken  as    part   of a   UK   Student
Fellowship at the Anglo-Australian Observatory.}

\section{Introduction}\label{introduction}
Recently, there has  been a rapid growth in  the number of extra-solar
planets identified (Vogt  et al. 2000; Marcy et  al. 2000).  Driven by
technological advances, most have  been found by the identification of
a `Doppler reflex', although  more novel techniques, such as planetary
occultation  (Charbonneau 2000), are  also proving  fruitful.  Collier
Cameron et al.   (1999) claim to have directly  detected the starlight
reflected  from the  planet orbiting  $\tau$ Bo\"{o}tis  (although see
Charbonneau  et al.   1999).  While  it now  appears that  this $\tau$
Bo\"{o}tis measure  has not been confirmed  in follow-up observations,
the identification  of planets due  to reflected starlight has  a firm
theoretical foundation (Sudarsky et al. 2000).

All these techniques focus on stars in the vicinity of the sun, over a
region  of several  tens of  parsecs.  At  larger  distances, however,
planets can  be detected orbiting the compact  objects responsible for
gravitationally   microlensing  stars  in   the  Galactic   Bulge  and
Magellanic Clouds via their perturbative influence on the microlensing
light curve  (Wambsganss 1997).  More recently, Graff  \& Gaudi (2000)
and Lewis  \& Ibata  (2000)(hereafter LI2000) have  considered instead
the detection of planets  orbiting the microlensed sources.  Using the
models for  `hot Jupiter' type planets  as a basis  for their studies,
these groups  demonstrated that the small fraction  of light reflected
from  the  planet  $(up~to  \times10^{-4}L_*)$  can  be  significantly
magnified, resulting  in an observable  $\sim1$ per cent  deviation in
the  microlensing  light  curve.   LI2000 also  demonstrated  how  the
monitoring of polarization through  a microlensing event can probe the
composition of the planetary atmosphere.

In  studying the  microlensing  light curve  of  planets orbiting  the
source  stars,   these  original  studies  assumed   that  the  planet
undergoing microlensing  can be represented by  a circularly symmetric
source.  This  situation only occurs  at opposition, and at  any other
point in its orbit the planet will display phases akin to the moon and
the  inferior planets.  This  paper considers  the influence  of these
deviations  from circular  symmetry on  the form  of  the microlensing
event with the goal of  examining their influence on the detectability
of  planets   via  gravitational  microlensing.   Section~\ref{method}
describes  the numerical approach  taken in  this paper,  and outlines
both  the  microlensing model  and  the  choice  of planetary  phases.
Section~\ref{results} describes  the results of  this study, including
an examination of how these results influence the detectability of the
planet, while Section~\ref{discos} presents the conclusions.

\begin{figure}
\centerline{\psfig{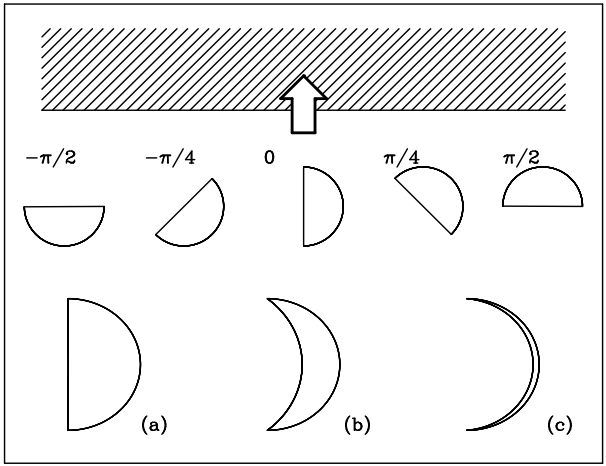}}
\caption{\label{planetphase}  The  planetary  phases and  orientations
considered in this paper. The cross-lines  and arrow at the top of the
figure  indicate the  caustic and  its direction  of motion,  with the
cross-lines   denoting   the   high   magnification  region   of   the
caustic.  Throughout   the  text  the   models  are  referred   to  as
$(i,\theta)$, where $i$ is the model  label (a,b or c) and $\theta$ is
the orientation.}
\end{figure}

\section{Method}\label{method}
\subsection{Microlensing Model}\label{micromodel}
Several  reviews  on  the  physical  aspects of  microlensing  in  the
Galactic Halo  have recently  appeared (e.g.  Paczy\'{n}ski  1986) and
this background  material will  not be reproduced  here.  As  with the
studies of  Graff \& Gaudi (2000)  and LI2000 this  paper assumes that
the planetary system is microlensed  by a binary system.  Such systems
can result in extended regions of strong magnification in the vicinity
of `fold caustics'.  Due to  their extended nature, such caustics have
proved useful probes of not only the surfaces of stars (Sasselov 1998;
Gaudi \& Gould 1999), including the detection of stellar spots (Han et
al. 2000),  but also the structure  at the heart of  quasars (Fluke et
al.  1999).  As  with Graff \& Gaudi (2000) and  LI2000, it is assumed
that  the  planet  is  swept   by  a  fold  caustic  during  a  binary
microlensing event.

The  magnification induced  as a  point source is  crossed   by a fold
caustic is given by
\begin{equation}
\mu(x) = {\frac{\kappa}{\sqrt{x-x_c}}} {H(x-x_c)} + \mu_0
\label{caustic}
\end{equation}
Here, \(x-x_{c}\) is  the distance from the source  to the caustic and
$H(x)$ is the Heaviside step function (Schneider et al 1992). The flux
factor  \(\kappa\)  represents  the  `strength'  of  the  caustic  and
\(\mu_{0}\) is the amplification of all other images of the source far
from the critical curve of the caustic (Chang 1984; Witt 1990; Wook et
al. 1998)

In general,  analytically calculating the light curve  of an arbitrary
source  as  it  is  microlensed  by a  fold  caustic  is  non-trivial.
Numerically, however,  the situation is much more  tractable. For this
study,  the magnification  distribution  due to  the  fold caustic  is
defined on a  grid of pixels.  The grid is chosen  such that the pixel
size is  much smaller than the  scale of the magnification  map or any
planetary structure. In considering pixels, Equation~\ref{caustic} can
written (Lewis \& Belle 1998)
\begin{equation}
\mu_{pix}  = 2{\frac{\kappa}{\Delta{x}}}{(\sqrt{x_k  +  \Delta {x}}  -
\sqrt{x_k})} + \mu_0\ .
\label{pixels}
\end{equation}
Here,  $\Delta{x}$  is the  extent  of the  pixel,  and  $x_k$ is  the
distance  of the  pixel to  the caustic.   In the  following  study we
assume $\mu_o=1$,  although we  will concentrate on  the magnification
due to the caustic.

\subsection{Planetary Phase}\label{phase}
Figure~\ref{planetphase}   summarizes   the   planetary   phases   and
orientations  employed in this  study.  The  lower panel  presents the
models for the phase; running from  (a) to (c) these go from a quarter
moon configuration to  an extreme crescent.  In the  upper part of the
Figure,  the  relative  orientation  of  the planetary  phase  to  the
incoming  caustic (denoted  by the  cross-lined region  and  arrow) is
presented.  Each model, therefore,  is labelled as $(i,\theta)$, where
$i$ represents the phase (a, b  or c) and $\theta$ is the orientation.
The crescent  is assumed to  be uniformly illuminated,  neglecting the
effects of  limb darkening. In  studies of stellar  microlensing, limb
darkening has  been found to  produce a variations of  several percent
during the  peak of a microlensing  event (e.g. Yock 2000)  and can be
considered a perturbation to the results presented here.

In comparing to previous work it  should be noted that due to symmetry
considerations,  the  magnification  of  model $(a,0)$  --  henceforth
referred to  as the  ``uniform, circular source''  -- is  identical to
that of  the circular source  of the same  radius. Note also  that, to
allow a comparison, in the  following the ``light curves'' present the
magnification over time, equivalent to assuming that the flux received
from each model  of the planet is the same,  irrespective of the phase
of the  planet.  The effect of  planetary phase on  the observed flux,
and  hence  detectability  of  the planetary  microlensing  event,  is
addressed in Section~\ref{detect}.

\begin{figure*}
\centerline{\psfig{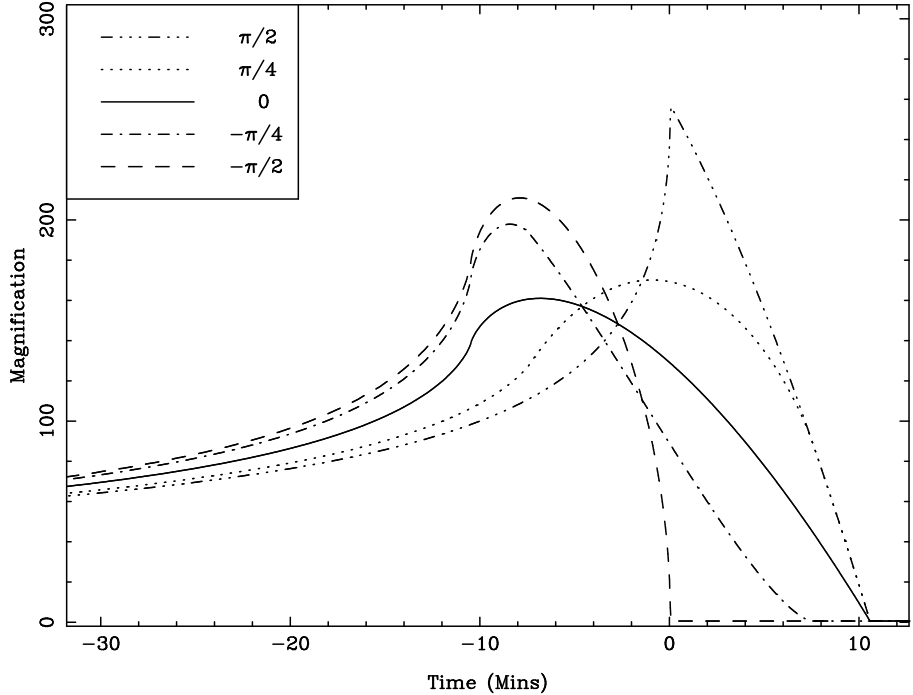}}
\caption{\label{half1}The light  curves for model  (a).  The different
line styles, described in the  key, refer to the fiducial orientations
displayed in Figure 1.  The  axes represent the magnification and time
scale for the microlensing of a hot Jupiter-like system located in the
Galactic bulge, allowing comparison with LI2000.}
\end{figure*}

\subsection{Numerical Approach}\label{nuapproach}
In  the  numerical approach  adopted,  each  source  was generated  as
depicted by models (a), (b)  and (c), in Figure \ref{planetphase}.  As
noted previously, the magnification pattern  was laid out on a grid of
pixels.   Similarly,  the  light  distribution over  the  planets,  as
illustrated in Figure~1, was also defined on a grid, taking account of
the orientation with respect to  the incoming caustic.  The grid scale
was  fine enough  to  ensure insignificant  numerical  errors, with  a
comparison to  analytic tests.  In  generating the lensed view  of the
source,  the planetary  profiled was  convolved with  the magnification
map, providing the light curve as  the source is swept by the caustic.
Note  that all  light curves  discussed in  Sections~\ref{quarter} and
\ref{Crescent}  are normalized  such  that the  unlensed  view of  the
source has unit flux.

\section{Results}\label{results}

\subsection{Quarter Moon Illumination: Model (a)}\label{quarter}
Figure~\ref{half1} presents the light curves for model (a) at the five
fiducial orientations described  in Section~\ref{phase}.  The axes are
for the model of LI2000  who considered a planet of radius 1.8$\times$
Jupiter located in the Galactic  Bulge, a value typical of hot Jupiter
systems.   The light  curves presented  in this  paper can  related to
those  of  differing  source   size,  caustic  strength  and  relative
velocities  through  simple  scaling  relations.   Firstly,  both  the
distances    and   caustic    strength    in   Equations~\ref{caustic}
and~\ref{pixels}  are   naturally  expressed  in   units  of  Einstein
radii. Given the relative velocities,  the crossing time of this scale
length is given by (e.g. LI2000);
\begin{equation}
t_E =
\frac{140}{v_{100}}\sqrt{\frac{M}{M_\odot}\frac{D_{os}}{8kpc}(1-d)d}
\ \ \ days
\label{timescale}
\end{equation}
where M is the mass of the lens (typically taken to be 1$M_\odot$) and
$v_{100}$  is the  perpendicular velocity  of the  lensing  objects in
units of  100$\kms$.  $D_{os}$ is the  distance to the  source in kpc,
while d is  the ratio of the  distance to the lens to  the distance to
the  source.  The  magnification can  be scaled  using the  results of
Chang (1984)  who demonstrated that when  a source of  radius $R_s$ is
swept  by a caustic  of strength  $\kappa$, it  will suffer  a maximum
magnification of
\begin{equation}
\mu_{max} = \frac{ f \kappa}{\sqrt{R_s}}
\label{peakmagnification}
\end{equation}
where $f$ is  the `form factor' which accounts  for the specific shape
and  brightness profile of  the source.   For a  uniformly illuminated
circular  disk, $f=1.39$.  

It is  immediately apparent that  rotating the source with  respect to
the  caustic radically  changes  the form  of  the microlensing  light
curves.  At  $(a,\pi/2)$, the light curve  is seen to  possess a rapid
rise,  sharply   turning  over  into  a   rapid  decline.   Similarly,
$(a,-\pi/2)$ also displays  a rapid rise, but a  more gentle turn over
into a  decline.  The intermediate  cases show similar  trends between
the two extreme orientations.

\begin{figure}
\centerline{\psfig{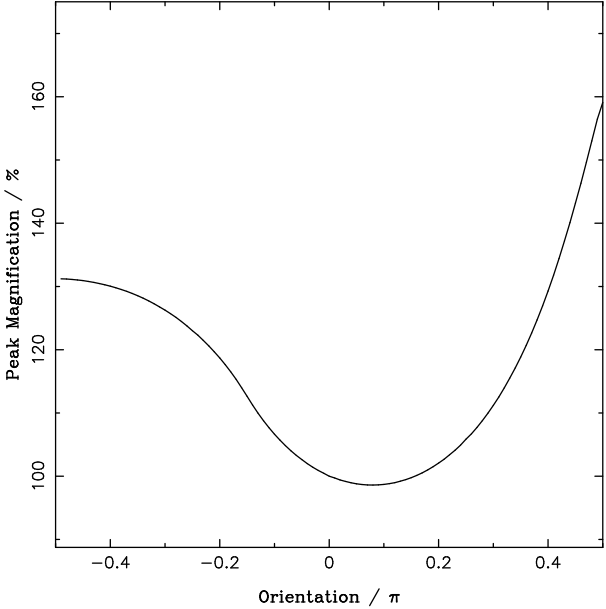}}
\caption{\label{half2}  The dependence  of peak  magnification  on the
orientation of the planet for  model (a).  The peak magnifications are
normalized to  the light curve  of the uniform, circular  source; this
will scale  to previous investigations on circular  sources, and allow
comparisons.}
\end{figure}

An examination of the  light curves in Figure~\ref{half1} reveals that
the  peak  magnification in  each  light  curve  also depends  on  the
planetary  orientation.  In  investigating this  further,  one hundred
additional light  curves at orientations between  $-\pi/2$ and $\pi/2$
were   generated.   Figure~\ref{half2}  presents   the  peak   of  the
magnification  of these  light curves  versus the  orientation  of the
planet  with  respect  to   the  caustic  (note  that  the  background
magnification, $\mu_o$,  has been subtracted from the  light curves in
this  analysis).  The  curve is  normalized with  respect to  the peak
magnification   of  model  $(a,0)$,   the  uniform,   circular  source
(equivalent  to a circular  source of  the same  radius).  Due  to the
scaling relations  outlined previously,  this curve is  insensitive to
the specific values of the caustic strength and planetary radius.

Interestingly, model  (a) suffers significantly  more magnification at
orientations of  $\pm\pi/2$ when compared  to an orientation of  0; in
the  case of  $\pi/2$ this  is  $\sim1.6\times$ that  of the  uniform,
circular source.   This result clearly demonstrates that  such a model
undergoes  a  minimum magnification  when  compared  to  model (a)  at
virtually  all  other  orientations.   The  degeneracy  between  model
$(a,0)$ and  a circularly symmetric  source of the same  radius allows
further    comparison   with    previous   results.     Hence,   using
Equation~\ref{peakmagnification}, the form factor $f$ for model (a) at
the  various  orientations  presented  in  this paper  can  be  simply
calculated by  scaling by  the values presented  in Figure~\ref{half2}
[using Figure~\ref{cres2} for models (b) and (c)].

\begin{figure*}
\centerline{\psfig{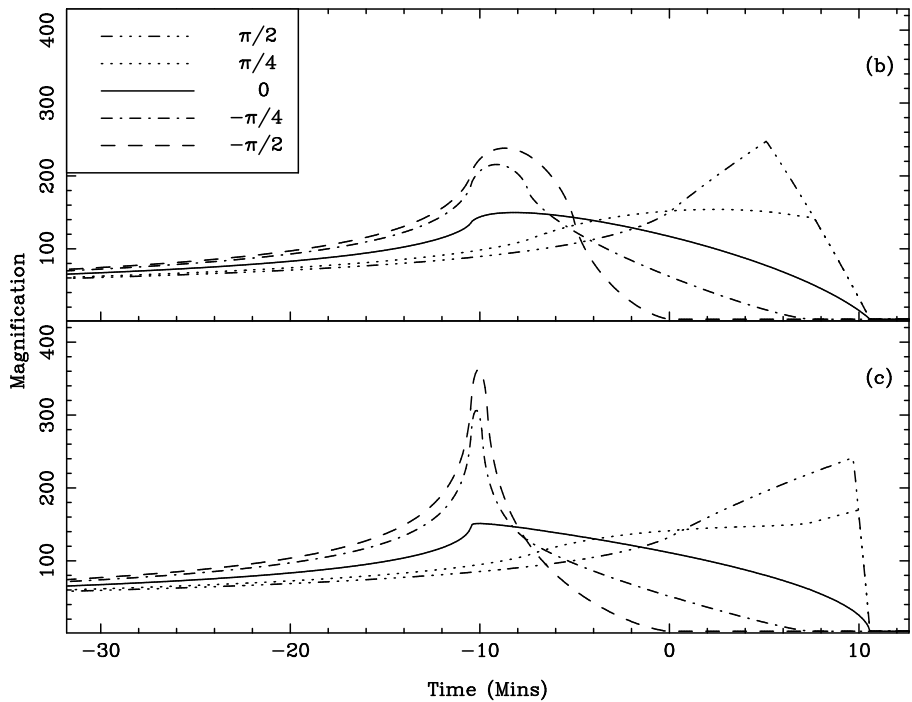}}
\caption{\label{cres1} As  Figure~2, but for  models  (b) and  (c). The
orientations in the key refer to the fiducial models denoted in Figure
1. In comparing to Figure~\ref{half1} it should be noted that the peak
magnification of  all models at  $\theta\sim0$ is virtually  the same.
As with  Figure~2, considering the microlensing of  a hot Jupiter-like
planet in the Galactic Bulge.}
\end{figure*}

\subsection{Crescent Illumination: Model (b) \& (c)}\label{Crescent}
Figure~\ref{cres1} presents  the light curves for models  (b) and (c).
In  comparing to  Figure~\ref{half1}  it  can be  seen  that the  peak
magnification of  all models at  $\theta\sim0$ is virtually  the same.
Both sets  of light curves possess  the same generic  features seen in
model  (a), although  as the  planet becomes  more  crescent-like, the
features  of  the light  curve  become  sharper.   This is  strikingly
apparent for model  (c) (lower panel), where the  light curve displays
an   almost   vertical   rise   in  $(c,-\pi/4)$   and   $(c,-\pi/2)$.
Correspondingly, $(c,\pi/4)$ and $(c,\pi/2)$ possess substantial peaks
in their light  curves, although these are accompanied  by more gentle
rises.

\begin{figure}
\centerline{\psfig{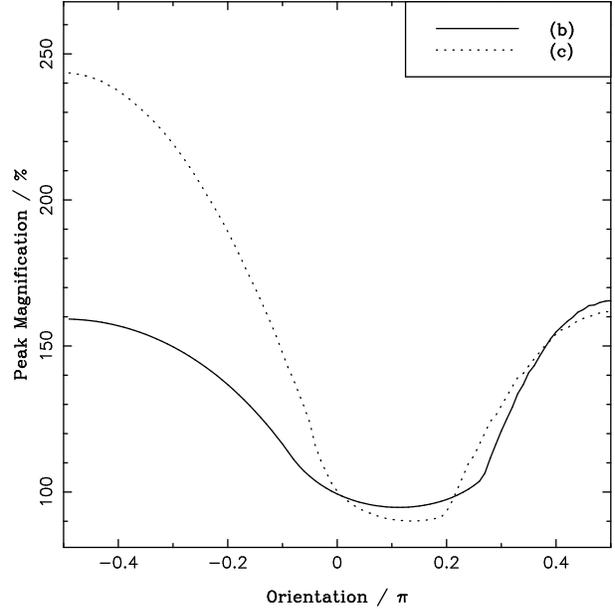}}
\caption{\label{cres2}  The dependence  of peak  magnification  on the
orientation of the planet, normalized by that of the uniform, circular
source.  Model  (b) is  denoted by a  solid line,  and model (c)  by a
dotted line.}
\end{figure}

Again,  the  peak  magnification   is  a  function  of  the  planetary
orientation with respect  to the caustic. As with  model (a), a larger
sample  of light  curves  for models  (b)  and (c)  were generated  to
investigate this dependence. The peak magnifications for these samples
are  presented  in  Figure~\ref{cres2}.   Interestingly, while  it  is
apparent that light curves for each model at $\theta=0$ are different,
they possess  very similar values  of peak magnification.   These have
been normalized  with the peak  magnification of the  uniform, circular
source.  As with model (a),  the light curves at $\theta\sim0$ possess
the lowest  value of peak magnification, with  higher values occurring
at  $\pm\pi/2$,  although  in  model  (b) the  extremes  of  the  peak
magnification  are at  $\sim1.6\times$ the  minimum value.   Model (c)
possesses  a similar  value at  $\pi/2$,  while at  $-\pi/2$ the  peak
magnification climbs to $\sim2.4\times$ the minimum value.

\subsection{Detectability}\label{detect}
The preceding  sections have discussed the  variation in magnification
as a  planet is swept  by a gravitational microlensing  caustic. While
this eases comparison with  previous models, being dependent only upon
the  size and  shape of  the source,  the observed  property  during a
microlensing  event is  flux.  For  a specific  planetary  system, the
fraction of  stellar light reflected  from a planet is  dependent upon
its  orbital phase.   As demonstrated  in the  previous  sections, the
details of the microlensing magnification are similarly dependent upon
the  planetary  orbital  phase,  and  it is  the  combination  of  the
reflected  planetary  flux   and  magnification  that  will  determine
detectability of the microlensing event.

\begin{figure*}
\centerline{\psfig{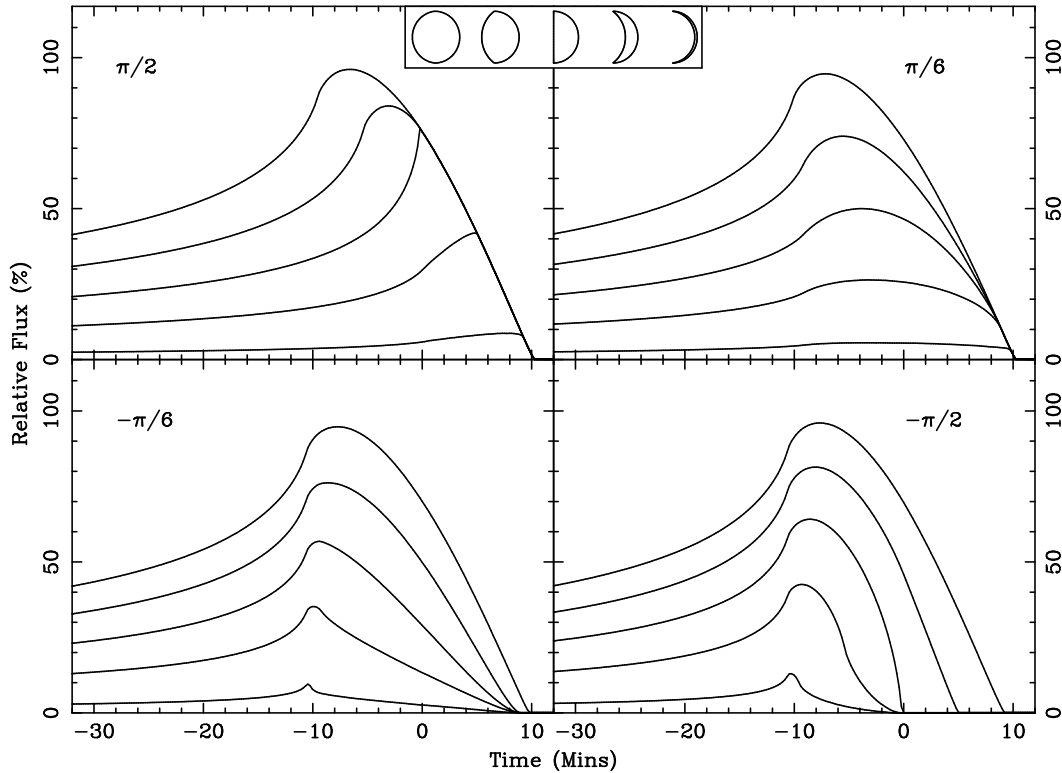}}
\caption{\label{relflux} A comparison of  the flux received during the
microlensing  of   a  particular   planet  at  differing   phases  and
orientations.   The orientation  of  the planet  with  respect to  the
incoming caustic  follows the convention  of Figure~\ref{planetphase},
and the phases under consideration are presented in the inset box. All
the light  curves are  normalized to the  peak flux received  when the
microlensed planet is full (i.e.  a circular source), which corresponds
to 100\%  on these plots.  All  panels display the  same feature, that
the amount  of flux received  falls monotonically as we  consider more
crescent-like source (left to right in the inset box), even though at
these phases the planet is more strongly magnified.}
\end{figure*}

With respect  to a  circular source, models  (a), (b) and  (c) reflect
50\%,  29\%  and  5\%  of  the stellar  light  respectively,  assuming
isotropic scattering.  Figure~\ref{relflux} presents the observed flux
for a  specific planet at several different  phases, including gibbous
phases (denoted in the inset  box).  Four orientations with respect to
the  sweeping  caustic  are   portrayed,  and  all  light  curves  are
normalised to the  flux of a microlensed circular  source, which peaks
at 100\%.  The same feature is  seen in all light  curves, namely that
the observed  flux drops as the  planetary phase changes  from full to
crescent,   even   though  these   latter   phases  undergo   stronger
magnification.

\begin{figure*}
\centerline{\psfig{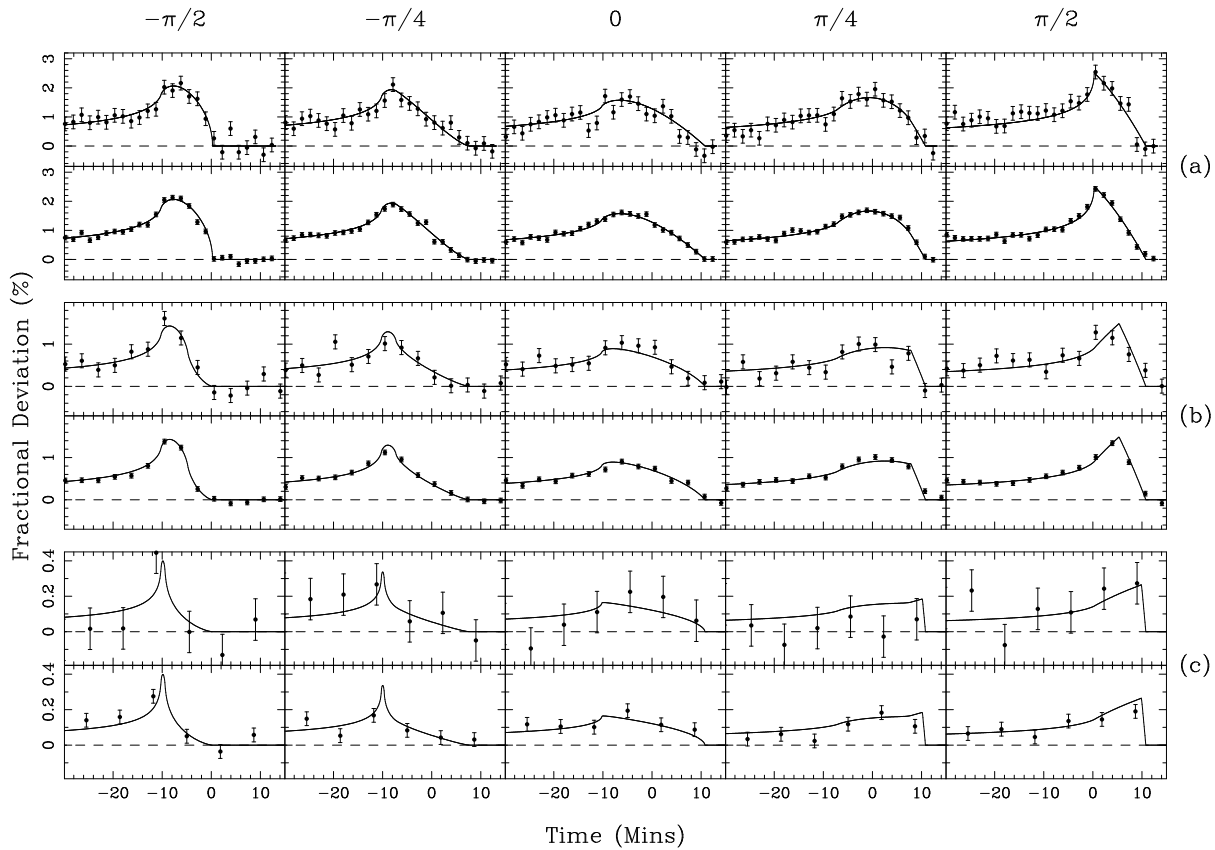}}
\caption{\label{noise}  Simulated observations  for  the light  curves
presented in Figures 2 \& 4.  For each model, presented vertically and
labelled (a),  (b) \&  (c), there  are two sets  of panels,  the upper
panel representing observations with  a 10-m telescope, while the lower
panel are observations with a 30-m telescope.}
\end{figure*}

Figure~\ref{noise}  presents simulated  observations  of light  curves
given in  Figures~2 and  4, assuming  the host star  is V=18,  with an
extinction  of 1.25mags (following  the model  of LI2000).   The three
models, (a),  (b) and (c)  are presented vertically, with  each column
representing the differing orientations.   For each model, two sets of
simulated observations are given, assuming a 10-m class (upper panels)
and 30-m class (lower panels),  using the planetary parameters used by
LI2000.  The  ordinate on each  panel represents the  fractional light
curve  deviation due  to the  presence of  the planet,  and  should be
compared to Figure~2  in LI2000.  Each telescope is  assumed to have a
10\%  total efficiency,  and the  data  points are  subject to  photon
counting errors  only.  The integration times for  both telescopes are
100, 200 and 400secs for model (a), (b) and (c) respectively.

The   10-m  class   telescope  observations   represent   the  current
astronomical  technology.  The fractional  deviation introduced  into a
light curve by  model (a), of order $\sim2\%$,  is clearly detectable,
with  the photometry  accurately revealing  the differing  light curve
profiles at  the different planetary orientations.  This  is also seen
with model  (b) which,  while possessing lower  signal-to-noise, still
allows recovery of  the detailed light curve profile.   For model (c),
however,  the  small  fraction  of  reflected  light  means  that  the
resultant  fractional  deviations  are  small, less  than  0.5\%.   As
illustrated in Figure~\ref{noise}, while fluctuations at this level are
potentially detectable, the details of  their profiles are lost in the
noise. Plans  are currently underway  to build the next  generation of
ground-based  telescope,  with   30-m  telescopes  entering  operation
between   2010-2015.   An   examination   of  the   lower  panels   in
Figure~\ref{noise}   demonstrate  that   with  the   advent   of  such
telescopes, the fractional deviation  introduced by a planet planetary
companion will  be readily  identifiable, even for  planets presenting
only a thin crescent.

It must  be remembered that the  above discussion is  for a particular
set  of  planetary  and  microlensing parameters,  such  as  planetary
radius, albedo, caustic strength and relative velocities, and that the
detectability  of a planetary  companion will  depend on  the specific
values of these for the system under consideration.

\section{Discussions \& Conclusions}\label{discos}
Recent  studies have  demonstrated that  when a  star in  the Galactic
Bulge or  Magellanic Clouds is magnified during  a microlensing event,
any light reflected from an  orbiting planet can also be significantly
magnified to  observable levels.  These previous works,  however, have
assumed  a simplified model  for the  distribution of  reflected light
from the  planet, namely a  circularly symmetric disk.   The reflected
light from any real planet, however, will display phases akin to those
seen with the moon and the inferior planets.

This paper has  presented a study on the effect  of planetary phase on
the degree to which the light  reflected from a planet is magnified as
it  is  swept  by   a  caustic  during  a  gravitational  microlensing
event.  Considering  simple  models  to  represent  various  planetary
phases, this  work has  demonstrated that these  non-circular profiles
result in strikingly different  light curves that are solely dependent
upon their  orientation with respect  to the sweeping  caustic.  While
this behaviour is different, depending upon the specifics of the light
distribution reflected by the planet,  all the models display the same
generic trends.

Equally interesting  is that the  peak magnification of a  light curve
for  a  model  of the  planet  is  also  strongly dependent  upon  the
planetary orientation with respect to the caustic.  In each model, the
planet was  substantially more magnified if  orientated at $\pm\pi/2$,
than at 0  (see Figure~\ref{planetphase}).  In the case  of the simple
`quarter moon'  (model a),  this resulted in  up to $\sim60$  per cent
more  magnification  than  when  compared  to  planets  orientated  at
$\sim0$. Due to scaling relations, Figures \ref{half2} and \ref{cres2}
are insensitive  to specific values of caustic  strength and planetary
radius.

These  trends are  also seen  in  models (b)  and (c)  which are  more
crescent-like, both  possessing peak magnification  distributions that
possess a minimum at $\theta\sim0$,  rising to $\pm\pi/2$. In the case
of  model  (c),   the  light  curve  at  $-\pi/2$   possesses  a  peak
magnification   exceeding   those   at   $\theta\sim0$  by   140   per
cent. Considering that model (a)  at $\theta=0$ is degenerate with the
circularly  symmetric   models  typically  employed   in  microlensing
studies, these results imply that previous results have underestimated
the  potential magnification  in  planetary microlensing  events by  a
factor of a few.

While these results are  encouraging, the detectability of a planetary
microlensing event  will depend on the  amount of flux  received by an
observer. As well  as the magnification, this will  depend also on the
amount of  stellar light that is  reflected by a planet,  a value that
decreases   as  a  planet   appears  more   crescent-like.   Simulated
observations presented  in this paper reveal that  the microlensing of
moderately crecsent-like planets (model b) are detectable with current
telescopes at a temporal sampling that allows the detailed recovery of
the underlying  profile.  With  more extreme crescents,  the planetary
event the fractional deviation  of ${\lta0.5\%}$ occurs too quickly to
be  effectively sampled by  photometric monitoring  with a  10-m class
telescope, although  proposed 30-m  class telescopes will  resolve the
fluctuations.   The  exact details  of  the  microlensing light  curve
deviations due to the presence  of a planetary companion depend upon a
number  of parameters,  although  the results  presented  here and  in
previous  studies (Graff  \&  Gaudi 2000;  LI2000)  suggest that  this
approach should successfully  uncover hot Jupiters at kiloparsec-scale
distances.

Due to condensed particulate matter in the atmospheres of planets, the
light they reflect  is expected to be polarized  (Seager et al. 2000),
although  the   degree  of   this  polarization  is   extremely  small
$(\sim10^{-5})$.   LI2000   considered  the  effect   of  microlensing
magnification on this polarization signature and found that it too can
be boosted to detectable levels,  although at a fraction of a percent,
the  measurement of  such polarizations  still  presents observational
difficulties. In light  of the results of the  study presented in this
paper,  we have  begun  an extensive  investigation  into the  further
effects  of  the  microlensing  of planets,  especially  the  expected
polarization signatures from crescent-like sources.

\section*{Acknowledgements}
The anonymous referee is thanked for comments that improved the paper.
GFL thanks Eric  Agol for discussions that initiated  this study.  CEA
thanks  the Anglo-Australian Observatory  for hospitality  and support
for the duration of her UK Student Fellowship.


\begin{thebibliography}{DUM}
%
\bibitem[Chang (1984)]{Chang1984}
Chang, K.
1984, A\&A. 130, 157
%
\bibitem[Charbonneau et al (1999)]{Charb1999}
Charbonneau, D., Noyes, R.W., Korzennik, S.G., Nisenson, P.,
Saurabh, J., Vogt, S.S. \& Kibrick, R.I.
1999, ApJ, 522, L145 
%
\bibitem[Charbonneau (2000)]{Charb2000}
Charbonneau, D., Brown, T.M., Latham, D.W. \& Mayor, M.
2000, ApJ, 529, L45
%
\bibitem[Collier Cameron, et al(1999)]{ColCametal1999}
Collier Cameron, A., Horne, K., Penny, A. \& James, D. 
1999, Nature, 402, 751
%
\bibitem[Graff & Gaudi (2000)]{Graff&Gaudi2000}
Graff, D. \& Gaudi, B.S.
2000, ApJ, 538, L133
%
\bibitem[Gaudi & Gould (1999)]{Gaudi&Gould1999}
Gaudi, B.S. \& Gould A.
1999, ApJ, 513, 619
%
\bibitem[Fluke & Webster (1999)]{Fluke1999}
Fluke, C.J. \& Webster, R.L.
1999, MNRAS, 302, 68
%
\bibitem[Han, et al (2000)]{Han2000}
Han, C., Park, S., Kim, H. \& Chang, K.
2000, MNRAS, 316, 665
%
\bibitem[Lewis & Belle (1998)]{Lewis&Belle1998}
Lewis, G.F. \& Belle, K.E.
1998, MNRAS, 297, 69
%
\bibitem[Lewis & Ibata (2000)]{Lewis&Ibata2000}
Lewis, G.F. \& Ibata, R.A.
2000, ApJ, 539, L63 (LI2000)
%
\bibitem[Marcy, et al (2000)]{Marcyetal2000}
Marcy, G.W., Butler, R.P. {\&} Vogt, S.S.
2000, ApJ, 536, L43
%
\bibitem[Paczy\'{n}ski (1986)]{Paczyki1986}
Paczy\'{n}ski, B.
1986, ARA\&A, 34, 410
%
\bibitem[Sasselov (1998)]{sasselov98} 
Sasselov, D.\ D.\ 
1998, ASP Conf.\ Ser.\ 154: Tenth Cambridge Workshop on 
Cool Stars, Stellar Systems, and the Sun, 10, 383 
%
\bibitem[Schneider, et al (1992)]{Schneideretal1992}
Schneider, P., Ehlers, J., {\&} Falco, E.E.
1992, Gravitational lenses (Berlin: Springer)
%
\bibitem[Seager, et al (2000)]{Seager2000}
Seager, S., Whitney, B.A. {\&} Sasselov, D.D.,
2000, ApJ, 540, 504
%
\bibitem[Sudarsky, et al (2000)]{Sudarsky2000}
Sudarsky, D., Burrows, A., {\&} Pinto, P.
2000, ApJ, 538, 885
%
\bibitem[Vogt, et al (2000)]{Vogt2000}
Vogt, S.S., Marcy, G.W., Butler, R.P. \& Apps, K.
2000, ApJ, 536, 902
%
\bibitem[Wambsganss (1997)]{Wambsganss1997}
Wambsganss, J.
1997, MNRAS, 284, 172
%
\bibitem[Witt (1990)]{Witt1990}
Witt, H.J.
1990, A\&A, 236, 311
%
\bibitem[Wook et al (1998)]{Wook1998}
Wook, L.D., Chang, K.Y., {\&} Joon, K.S. 
1998, J. Korean Astron. Soc., 31, 27
%
\bibitem[Yock (2000)]{Wo}
Yock, P.
2000, PASA, 17, 35
%
\end{thebibliography}
\end{document}